# Energy efficiency studies for dual-grating dielectric laser-driven accelerators

Y. Wei [1,2,*], M. Ibison [1,2], J. Resta-Lopez [1,2], C. P. Welsch [1,2], R. Ischebeck [3], S. Jamison [4], G. Xia [1,5], M. Dehler [3], E. Prat [3], J. D. A. Smith [6]

[1]Cockcroft Institute, Sci-Tech Daresbury, Warrington, WA4 4AD, United Kingdom
[2]Department of Physics, University of Liverpool, Liverpool, L69 7ZE, United Kingdom
[3]Paul Scherrer Institute, Villigen, 5232, Switzerland
[4]Accelerator Science and Technology Centre, Sci-Tech Daresbury, Warrington, WA4 4AD, United Kingdom
[5]School of Physics and Astronomy, University of Manchester, Manchester, M13 9PL, United Kingdom
[6]Tech-X UK Ltd, Sci-Tech Daresbury, Warrington, WA4 4AD, United Kingdom

Dielectric laser-driven accelerators (DLAs) can provide high accelerating gradients in the GV/m range due to their having higher breakdown thresholds than metals, which opens the way for the miniaturization of the next generation of particle accelerator facilities. Two kinds of scheme, the addition of a Bragg reflector and the use of pulse-front-tilted (PFT) laser illumination, have been studied separately to improve the energy efficiency for dual-grating DLAs. The Bragg reflector enhances the accelerating gradient of the structure, while the PFT increases the effective interaction length. In this paper, we investigate numerically the advantages of using the two schemes in conjunction. Our calculations show that, for a 100-period structure with a period of 2 μm, such a design effectively increases the energy gain by more than 100 % when compared to employing the Bragg reflector with a normal laser, and by about 50 % when using standard structures with a PFT laser. A total energy gain of as much as 2.6 MeV can be obtained for a PFT laser beam when illuminating a 2000-period dual-grating structure with a Bragg reflector.

## 1. Introduction

Photonic microstructures [1-8] have been observed to transfer the energy of a fiber laser to electrons by sustaining an accelerating field to interact with electrons in vacuum. This has enabled the development of dielectric laser-driven accelerators (DLAs). These microstructures made of dielectrics can offer high accelerating gradients in the range of GV/m, and have the potential to dramatically reduce the size and cost of future particle accelerators. To date, proof-of-principle experiments have demonstrated accelerating gradients of 300 MV/m [9], 690 MV/m [10], and 1.8 GV/m [11] for relativistic electrons, and gradients of 25 MV/m [12], 220 MV/m [13] and 370 MV/m [14] for non-relativistic electrons. These demonstrations pave the way for implementing an on-chip particle accelerator in the future.

The dual-grating structures proposed by Plettner *et al.* [1] are of particular interest because they have a simpler geometry than other types of DLA, which reduces the complexity and expense of the fabrication process. When they are illuminated by a uniform plane wave, longitudinal electric fields $E(z,t)$ are generated in the vacuum channel center, where the electrons travel and are accelerated. Here, $z$ is the longitudinal position of the electrons in the vacuum channel at a time $t$. The accelerating gradient $G_0$ is defined as the average electric field that an electron experiences in one grating period:

---

* Corresponding author: yelong.wei@cockcroft.ac.uk.

$$G_0 = \frac{1}{\lambda_p} \int_0^{\lambda_p} E(z,t)\, dz, \quad (1)$$

where $\lambda_p$ is the grating period. When the phase slippage is not into account, the total energy gain within the entire interaction length $L_{int}$ can be expressed as

$$\Delta E = G_0 L_{int}. \quad (2)$$

When a Gaussian laser beam with a temporal profile $s(t, \tau_0)$, and a spatial profile $g(z, w_z)$ is used to illuminate a dual-grating structure with a length $LZ$, the electrons' energy gain is calculated to be

$$\Delta E = \int_{-0.5LZ}^{0.5LZ} q G_p s(t, \tau_0) g(z, w_z)\, dz, \quad (3)$$

where $\tau_0$ is the laser full-width at half-maximum (FWHM) duration, $w_z$ is the laser transverse RMS waist radius, $q$ is the charge on a single electron, and $G_p$ is the accelerating gradient. Since the structure is not powered uniformly, the accelerating gradient $G_p$ is the peak gradient. Here, we define $z = 0$ as corresponding to the longitudinal center of a dual-grating structure and the center of the laser transverse profile. Equation (3) shows that the energy gain depends on the laser FWHM duration $\tau_0$, waist radius $w_z$, structure length $LZ$, and gradient $G_p$.

The ratio of the energy gain $\Delta E$ and the product of the incident laser field $E_0$ and $q$ is defined as the energy efficiency $\eta$, which can be used to evaluate the accelerating performance:

$$\eta = \Delta E / (q E_0). \quad (4)$$

A normally-incident laser illumination has a temporal profile $s(t, \tau_0) = e^{-2\ln 2 \left(\frac{t}{\tau_0}\right)^2}$ and a spatial profile $g(z, w_z) = e^{-\left(\frac{z}{w_z}\right)^2}$. For this case, the energy gain can be expressed as:

$$\Delta E_n = \int_{-0.5LZ}^{0.5LZ} q G_p e^{-\left(\frac{z}{w_{int}}\right)^2} dz. \quad (5)$$

where $w_{int} = \left(\frac{1}{w_z^2} + \frac{2\ln 2}{(\beta c \tau_0)^2}\right)^{-0.5}$ is the characteristic interaction length, as described in Refs. [15, 16], $\beta = v/c$, where $c$ is the speed of light.

A PFT laser illumination corresponds to a temporal profile $s(t, \tau_0) = e^{-2\ln 2 \left(\frac{t-pz}{\tau_0}\right)^2}$ and a spatial profile $g(z, w_z) = e^{-\left(\frac{z}{w_z' \cos\gamma}\right)^2}$, where $p = \frac{dt}{dz}$ is the PFT factor which is defined by the derivative of the pulse-front arrival time with respect to $z$, $w_z'$ is the tilt waist radius which can be obtained through a diffraction grating and a imaging system, and $\gamma$ is the tilt angle, as shown in Fig. 1. As described in Ref. [17], we achieve $t - pz = 0$ when electrons move at a speed $v = \beta c$ which matches with tilt angle $\gamma$. In this case, the energy gain is

$$\Delta E_{PFT} = \int_{-0.5LZ}^{0.5LZ} q G_P e^{-\left(\frac{z}{w_{int}'}\right)^2} dz, \quad (6)$$

where $w_{int}' = w_z' \cos\gamma$ is also the characteristic interaction length, as described in Ref. [17]. Compared to Eq. (5), Eq. (6) shows that the energy gain is related only to tilt waist radius $w_z'$ and tilt angle $\gamma$.

In order to increase the energy efficiency, geometry optimization studies have been carried out for dual-grating DLAs [1-4, 18]. In addition to this, two new kinds of scheme were proposed [1,19] and studied [17,20] separately to improve the energy efficiency for dual-grating DLAs. One scheme was to introduce a Bragg reflector [19], consisting of many layers of dielectric, into bare dual-grating structures. This scheme has been observed numerically in Ref. [20] to boost the accelerating field in the channel and thereby increase the energy efficiency by more than 70% compared to bare dual-gratings. Pulse-front-tilt (PFT) operation for the laser beam was proposed in Ref. [1] as a second scheme to extend the interaction length. This scheme would effectively increase the energy efficiency by more than 100 % compared to that of normally-incident laser illumination, as studied in Ref. [17].

Combining these two schemes, we present in this paper detailed numerical studies of a dual-grating structure with a Bragg reflector driven by a PFT laser beam, which can generate an energy efficiency higher than either scheme separately. As shown in Fig. 1, a PFT laser beam is introduced to interact with an electron bunch in a dual-grating structure with a 7-layer Bragg reflector. The reflector reflects back laser power to enhance the accelerating field in the channel, thereby increasing the accelerating gradient for electrons and greatly improving the energy efficiency. It should be noted that a 7-layer Bragg reflector is chosen, following an optimization study in Ref. [20]. Quartz [1, 21-22], with a refractive index $n = 1.5$, is chosen as the dielectric material for the structure and the Bragg reflector due to its high level of transparency in the optical frequency range, and its high field damage threshold and high thermal conductivity. The tilt angle $\gamma$ can be chosen to overlap an electron bunch synchronously with the laser pulse envelope, so that the extended interaction length allows electrons to gain the largest possible energy. Section 2 presents detailed 2D particle-in-cell simulations in which a PFT laser beam is introduced into a 100-period structure to interact with a 50 MeV electron bunch. Section 3 investigates the dependencies of the energy gain on the incident laser waist radius and the number of structure periods. Finally, the potential challenges are discussed.

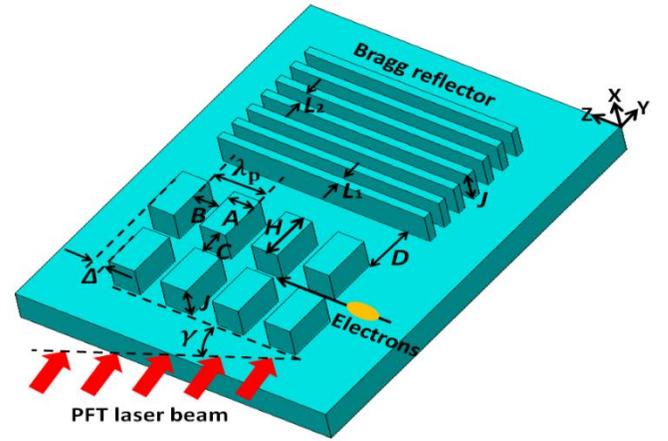

Fig. 1. Schematic of a dual-grating structure with a 7-layer Bragg reflector driven by a PFT laser beam with a tilt angle $\gamma$. $\lambda_p$, $A$, $B$, $C$, $D$, $H$, $J$, $L_1$, $L_2$, and $\Delta$ represent grating period, pillar width, pillar trench, vacuum channel gap, distance between dual-gratings and Bragg reflector, pillar height, vertical size, dielectric-layer thickness, vacuum-layer thickness, and longitudinal shift, respectively. The condition $A + B = \lambda_p$ is selected for all simulations to ensure synchronicity.

## 2. Particle-in-cell simulations

The particle-in-cell code VSim [23] is used to investigate the interaction between a front-tilted laser pulse and a Gaussian electron bunch, in a 100-period dual-grating structure with a 7-layer Bragg reflector having geometries: $A = B = 0.50\lambda_p$, $C = 0.50\lambda_p$, $H = \lambda_p$, $\Delta = 0$ nm, $D = 0.80\lambda_p$, $\lambda_p = 2.0$ µm, and total length $LZ = 200$ µm. These geometries have been studied in Ref. [20]. The electron bunch employed in our simulations has a mean energy of 50 MeV, bunch charge of 0.1 pC, RMS length of 9 µm, RMS radius of 10 µm, normalised emittance of 0.2 mm·mrad, and energy spread of 0.03%. Such an electron bunch can be achieved at the future Compact Linear Accelerator for Research and Applications (CLARA) [24] or the Advanced Superconducting Test Accelerator (ASTA) at Fermilab [25].

An incident Gaussian laser pulse with $\lambda_0 = 2.0$ µm wavelength, $E = 2.1$ µJ pulse energy, $\tau_0 = 100$ fs pulse duration, and $w_z = 50$ µm waist radius would generate a peak input field $E_0 = 2.0$ GV/m. When such a laser pulse is used for normal illumination, the maximum field generated in the structure area is still below the damage threshold for quartz structures [1,21-22]. The calculated interaction length is $w_{\text{int}} = \left(\frac{1}{w_z^2} + \frac{2\ln 2}{(\beta c\tau_0)^2}\right)^{-0.5}$ = 22.7 μm, which is much smaller than $LZ$ = 200 μm. Using Eq. (5) with $LZ$ = 200 μm and a peak gradient of $G_P$ = 1.0 GV/m we find a maximum energy gain of $\Delta E_n \approx qG_P\sqrt{\pi}w_{\text{int}}$ = 40 keV. This is used to calculate the loaded gradient for subsequent analysis of normally-incident laser illumination.

The same laser parameters are used for the optical system described in Ref. [17], to generate a front-tilted pulse with an ultrashort pulse duration $\tau_0$ = 100 fs and a tilt angle $\gamma = 45^0$. As calculated in Ref. [17], an interaction length of $w'_{\text{int}}$ = 51 μm is obtained for an incident laser waist radius $w_z$ = 50 μm. When such a front-tilted pulse propagates through the structure to interact with the electron bunch within it, the maximum energy gain is calculated to be $\Delta E_{\text{PFT}} \approx qG_p\sqrt{\pi}w'_{\text{int}}$ = 90 keV using Eq. (6) with $LZ$ = 200 μm and $G_p$ = 1.0 GV/m. This is also used to calculate the loaded gradient for subsequent simulations of PFT laser illumination.

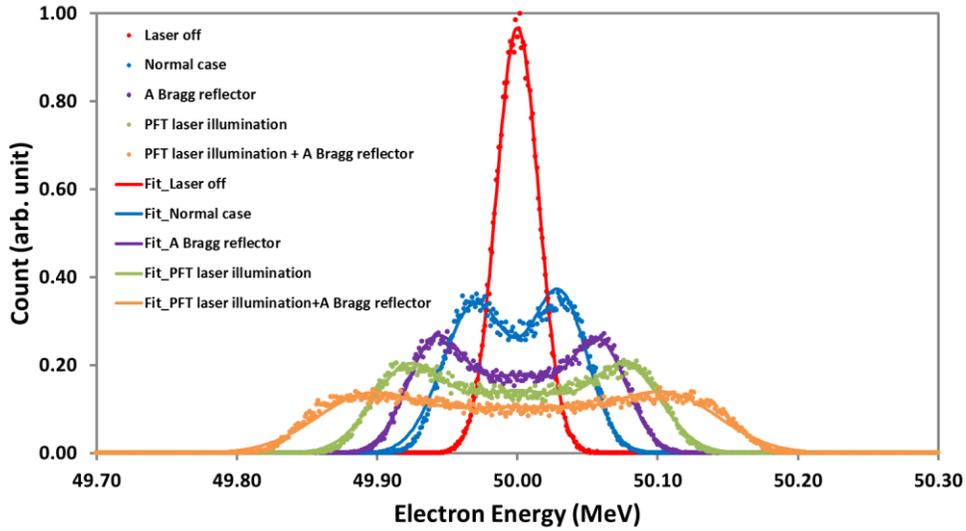

Fig. 2. Bunch energy distribution for the cases of laser-off (red dots and fit line), laser-on with a normal laser pulse for bare dual-gratings (blue dots and fit line), laser-on with a normal laser pulse for dual-gratings with a Bragg reflector (purple dots and fit line), laser-on with a front-tilted laser pulse for bare dual-gratings (green dots and fit line), and laser-on with a front-tilted laser pulse for dual-gratings with a Bragg reflector (yellow dots and fit line).

In the particle-in-cell simulations, the mesh size is set to 10 nm (z) × 20 nm (y) so that the results are convergent. We use 500,000 macro particles for tracking, close to the number of electrons for a bunch charge of 0.1 pC. The bunch transverse size is much bigger than the vacuum channel gap of $C$ = 1.0 μm, so only a small fraction of the electrons traverses the vacuum channel of the structure - it is calculated that about 4% of the 50 MeV bunch is transmitted through the vacuum channel gap of 1.0 μm. Those electrons travelling through the quartz structure suffer significant energy loss due to collisional straggling [26] in the dielectric material. We use only the electrons modulated by the laser field in the vacuum channel for subsequent analysis.

Figure 2 shows the results of the particle-in-cell simulations for the different configurations. Using the Gaussian fits to these energy spectra, the maximum energy gain is calculated from the difference between the abscissa of the half-width at half-maximum (HWHM) point for a laser-on spectrum and a laser-off spectrum [9]. The maximum energy gain is $\Delta E_1$ = 35 keV for normal laser illumination on bare dual-gratings, $\Delta E_2 = 63$ keV for normal laser illumination on dual-gratings with a Bragg reflector, $\Delta E_3 = 88$ keV for PFT laser illumination on bare dual-gratings, and $\Delta E_4 = 131$ keV for PFT laser illumination on dual-gratings with a Bragg reflector. This corresponds to maximum loaded gradients of 0.88 GV/m, 1.58 GV/m, 0.98 GV/m, and 1.46 GV/m, respectively. As expected, while PFT laser illumination has a similar loaded gradient to the normal one, a Bragg reflector boosts the loaded gradient, and hence the energy efficiency, for dual-grating DLAs. The loaded accelerating gradients are increased by 80 % and 49 % for normal and PFT laser illumination respectively, when a Bragg reflector is added, for bare dual-grating structures. For dual-gratings with a Bragg reflector driven by a PFT laser beam, the energy gain is increased by 274 % as compared to normally-incident laser illumination on bare dual-gratings.

## 3. Energy gain analysis

In this section, we study further the relationship between the energy gain $\Delta E_4$ when a PFT laser and a Bragg reflector are used and the waist radius $w_z$ of the incident laser beam. As observed in Ref. [17], the interaction length $w'_{\text{int}}$ increases linearly with the waist radius $w_z$ of the incident laser beam. Using Eq. (6) with a loaded gradient $G_p = 1.46$ GV/m, we can calculate analytically the energy gain $\Delta E_4$ with the incident laser waist radius $w_z$. Figure 3 shows the energy gain $\Delta E_4$ from analytical and particle-in-cell calculations with variable laser waist radii $w_z$; very good agreement is found between the two approaches. As shown in Fig. 3, the energy gain gradually increases with laser waist radius, and it reaches saturation when the radius is larger than 400 μm. For a 100-period dual-grating structure with a Bragg reflector driven by a PFT laser, the maximum energy gain is calculated to be 290 keV.

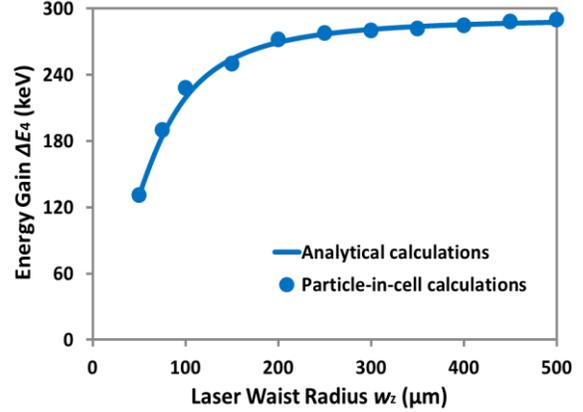

Fig. 3. Energy gain from analytical (solid line) and particle-in-cell (dots) calculations as a function of the laser waist radius, for a 100-period dual-grating structure with a Bragg reflector driven by a PFT laser: $E_0 = 2$ GV/m, $\tau_0 = 100$ fs, and $\gamma = 45^0$.

It should be noted that a 100-period structure with a length $LZ = 200$ μm is used for our particle-in-cell simulations due to limited computing resources. As seen in Ref. [17], a larger number of grating periods results in a higher energy gain for a bare dual-grating structure. This also applies to dual-gratings with a Bragg reflector. Using Eq. (6) with a loaded gradient $G_p = 1.46$ GV/m, the energy gain $\Delta E_4$ with variable structure lengths $LZ$ and laser waist radii $w_z$ can be calculated analytically. Figure 4 shows the analytically-calculated energy gain, which gradually increases to reach saturation with the laser waist radius $w_z$ from 50 μm to 1000 μm for different periods of the structures. For a dual-grating structure of more than 2000 periods, corresponding to a length $LZ = 4000$ μm, the energy gain increases linearly with laser waist radius from 50 μm to 1000 μm. In this case, a laser waist radius of 1000 μm would generate a maximum energy gain of 2.6 MeV.

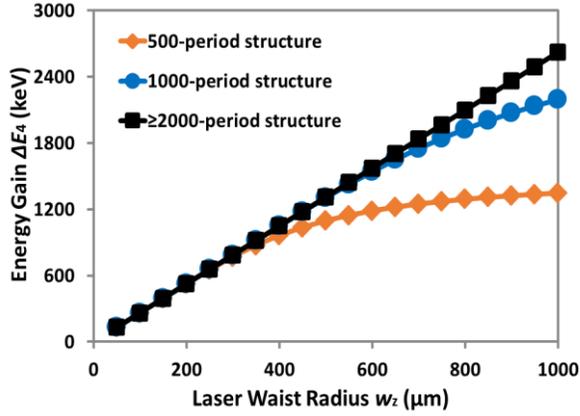

Fig. 4. Analytically-calculated energy gain with increasing laser waist radius, for different periods of a dual-grating structure with a Bragg reflector driven by a PFT laser: $E_0 = 2$ GV/m, $\tau_0 = 100$ fs, and $\gamma = 45^0$.

## 4. Conclusion

In conclusion, we have studied numerically the energy efficiency for dual-gratings with a 7-layer Bragg reflector driven by a PFT laser beam. When a PFT laser beam with a peak field of 2.0 GV/m is used to illuminate such a structure with 100 periods, the energy gain generated is increased by about three times as compared to normal laser illumination on bare dual-gratings.

We have also verified that the energy gain for dual-gratings with a Bragg reflector driven by a PFT laser beam depends strongly on the incident laser waist radius and the number of structure periods. For a 100-period dual-grating structure, the maximum energy gain is calculated to be 290 keV for a laser waist radius of about 500 μm. For a structure with more than 2000 periods, the energy gain increases up to 2.6 MeV for an incident laser with a waist radius of 1000 μm.

In short, the combination of a Bragg reflector and PFT laser illumination can be employed to improve the energy efficiency for any DLA structure. However, realistic fabrication and experimental studies are still required, to pave the way for implementing the proposed concept for such an integrated nano-structure driven by a PFT laser beam.

## Acknowledgements

We would like to thank Dr. Eugenio Ferrari and Dr. Nicole Hiller for many useful discussions. This work was supported by the EU under Grant Agreement 289191 and STFC Cockcroft Institute core grant No.ST/G008248/1.